\documentclass[letterpaper, 10 pt, conference]{ieeeconf} 
\IEEEoverridecommandlockouts  
\usepackage{graphics}
\usepackage{graphicx}
\usepackage{epsfig}
\usepackage{multirow}
\usepackage{amssymb}
\usepackage{subcaption}
\usepackage{amsmath}
\usepackage{booktabs}
\usepackage{longtable}
\usepackage{stackengine}
\usepackage{xfrac}
\usepackage{tabularx}
\usepackage[dvipsnames]{xcolor,colortbl}
\usepackage[clock]{ifsym} 
\usepackage{float} 
\usepackage[normalem]{ulem}
\usepackage{bm} 

\usepackage{soul}

\usepackage{cite}
\usepackage{hyperref}
\usepackage{url}

\usepackage{algorithm}      
\usepackage{algorithmic}    

\usepackage{epstopdf} 
\usepackage{makecell} 
\usepackage[table]{xcolor} 
\definecolor{lightgray}{gray}{0.9}
\definecolor{lightblue}{rgb}{0.85,0.92,1.0}
\definecolor{lightred}{rgb}{1.0,0.85,0.85}

\begin{document}

\graphicspath{{Figures_R0/}}
\allowdisplaybreaks

\title{Adaptive Input Shaper Design for Unknown Second-Order Systems with Real-Time Parameter Estimation}
\author{Nyi Nyi Aung, Bradley Wight, and Adrian Stein
\thanks{N. N. Aung, B. Wight, and A. Stein are with the Department of Mechanical and Industrial Engineering, Louisiana State University, LA 70803, USA. {\tt\small(email: \{naung1,bwight3,astein\}@lsu.edu)}.}}

\maketitle

\begin{abstract}
We propose a feedforward input-shaping framework with online parameter estimation for unknown second-order systems. The proposed approach eliminates the need for prior knowledge of system parameters when designing input shaping for precise switching times by incorporating online estimation for a black-box system. The adaptive input shaping scheme accounts for the system's periodic switching behavior and enables reference shaping even when initial switching instants are missed. The proposed framework is evaluated in simulation and is intended for vibration suppression in motion control applications such as gantry cranes and 3D printer headers.
\end{abstract}

\IEEEpeerreviewmaketitle

\vspace{-0.1in}
\section{INTRODUCTION}
\label{sec:introduction}
Input shaping is a feedforward control method that prefilters the reference command to suppress residual vibrations in flexible or underdamped systems. This is achieved by convolving the desired input with time delay filter (TDF) whose impulse amplitudes and delays are selected to reduce excitation of specific modes. The method is versatile for diverse applications ranging from cranes to robotics. A comprehensive frequency-domain treatment of input shaper design was presented by Singh~\cite{singh_optimal_2010}, forming the basis for numerous extensions. Subsequent work addressed robustness to modeling uncertainties through convex optimization–based finite impulse response (FIR) design~\cite{stein_convex_2024} and adaptive arbitrary-time-delay input shaping~\cite{jiang_vibration_2025}. For gantry cranes, input shaping has been applied to inertial payload handling~\cite{stein_input_2022} and integrated with minimum-time control under velocity constraints~\cite{stein_minimum_2023}. Shapley Effects have also been used to assess and improve robustness~\cite{stein_shapley_2025}, while other work has demonstrated robust swing suppression for overhead cranes~\cite{awi_robust_2024}. Wang et al.~\cite{wang_time_2023} addressed trajectory distortion in multi-axis shaping by establishing a mapping between the time-parameter domains of pre- and post-shaped trajectories for improved contour error estimation.

Data-driven and AI-based methods are also advancing the field. Work~\cite{grazioso_input_2017} used artificial neural networks to obtain closed-form expressions for impulse amplitudes and locations in real time. Work~\cite{yang_vibration_2024} combined extended Kalman filtering with residual neural networks for robust shaping, while work~\cite{ding_deep_2025} applied deep reinforcement learning to double-pendulum crane trajectory planning and work~\cite{yang_data_2024} provided a broader survey of data-driven vibration control methods. Hybrid and alternative vibration suppression strategies have also been explored. These include hybrid fuzzy-logic and input shaping control~\cite{alhassan_hybrid_2024}, passivity-based crane control~\cite{shao_interconnection_2024}, adaptive anti-swing methods~\cite{li_adaptive_2024}, and block backstepping–based trajectory tracking~\cite{li_novel_2025}. In more complex dynamics, input shaping has been generalized to flexible systems with nonlinear springs~\cite{moonmangmee_generalized_2025} and applied to two-mass drive systems~\cite{chen_vibration_2023}. Predictive shaping has also been implemented in embedded systems using model predictive control frameworks~\cite{van_den_broeck_embedded_2010}.

A topic which has been underexplored is real-time estimation with input shaping. Early contributions in time-varying shaping for vibration reduction in industrial robots~\cite{chang_time-varying_2005} and real-time shaping for precise XY stages~\cite{park_study_2008}. Advances in sensing and computation have enabled new real-time estimation approaches. Work~\cite{stein_aruco_2024} demonstrated the use of ArUco fiducial markers to initiate motion with small, arbitrary step commands, enabling online estimation of modal parameters and apply input shaping on the fly. It should be mentioned that there is a large potential for fiducial marker navigation in the unmanned ground and aerial vehicles research~\cite{fagundes-junior_uavugv_2025}.

Although these developments have greatly expanded the scope of input shaping, a persistent challenge remains: accurate parameter estimation within a predefined time window can limit real-time shaping effectiveness. This work addresses this limitation by introducing a closed-form, real-time input shaping framework for arbitrary undamped second-order systems with unknown parameters. Compared to work~\cite{stein_aruco_2024} where the closed form expression was dependent on the estimation time being smaller than the switch time, this manuscript is not limited by the estimation time. The proposed method initiates motion with a small step to estimate system parameters online, then automatically determines the optimal shaper activation time based on amplitude scaling, removing the need for fixed-duration estimation and enabling flexible, adaptive deployment for the user. To the best of the authors literature review this is the first method addressing this issue.

All models and implementations used in this work are made publicly available at \mbox{\texttt{https://github.com/NyiNyi-14/A-TDF.git}}. The methodology of the research is detailed in Section~\ref{sec:methodology}, followed by the presentation of numerical results in Section~\ref{sec:results}. Finally, Section~\ref{sec:conclusion} concludes the paper with a summary of the key findings.
%
%
%
\section{METHODOLOGY}
\label{sec:methodology}
\begin{figure}
    \centering
    \includegraphics[clip, trim= 20 700 320 0,width=1\linewidth]{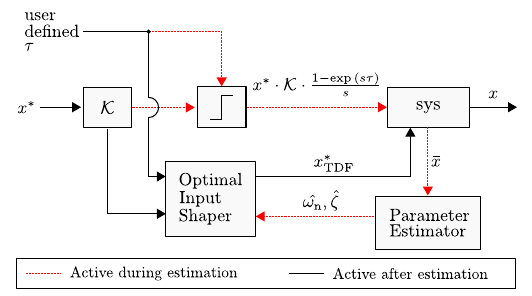}
    \caption{Feedforward control scheme with parameter estimation and optimal input shaping for vibration suppression.}
    \label{fig:block}
\end{figure}

Two main methodologies are employed: parameter estimation and optimal input shaper design. As shown in Fig.~\ref{fig:block}, the red dotted lines are active only during the estimation phase of user-defined duration $\tau$. In this phase, a scaled portion of the reference, $\mathcal{K} \cdot x^*$, where $\mathcal{K}$ denotes the reference fraction, is applied as a small step input:  
$x^* \cdot \mathcal{K} \cdot \frac{1 - e^{-s \tau}}{s},$
with $\mathcal{K}=0.01$ in this work. From the resulting response $\bar{x}$, the parameter estimation block calculates the natural frequency $\omega_\mathrm{n}$ and damping ratio $\zeta$ of the second-order system. The estimated parameters, $\hat{\zeta}$ and $\hat{\omega}_\mathrm{n}$, are then used to design the optimal input shaper via the TDF, generally expressed as~\cite{singh_optimal_2010}:
\begin{equation} \label{eq:is_nom}
G(s) = A_0 + A_1 e^{-sT},
\end{equation}
where $T$ is the switch time, and $A_0$, $A_1$ are the magnitudes at $t=0$ and $t=T$. When parameter estimation time is considered, \eqref{eq:is_nom} is adapted accordingly using either fixed- or arbitrary-time approaches, as detailed in the following sections. The resulting TDF reference $x^*_\mathrm{TDF}$ is then applied to the open-loop system to achieve vibration suppression.
%
%
%
\subsection{Parameter Estimation in Feedforward Control}
\label{sec:estimator}
Consider the black-box second-order system as:
\begin{equation}
    P(s) = \frac{\omega_\mathrm{n}^2}{s^2 + 2\zeta\omega_\mathrm{n}s + \omega_\mathrm{n}^2},
\end{equation}
where parameter estimation is performed according to Algorithm~\ref{para_algo}. In this context, $\frac{\mathrm{d}\bar{x}(t)}{\mathrm{d}t}$ denotes the slope of the response, $M_p$ the overshoot, $T_\mathrm{s}$ the settling time, $\Delta \bar{x}(t_{\mathrm{p}k})$ the difference between two successive peaks, $\Delta t$ the time interval between peaks, and $T_\mathrm{Mp}$ the time of peak overshoot. The algorithm takes $\bar{x}$ as input, determines whether the system is damped or undamped from its transient behavior, and computes $\hat{\zeta}$ and $\hat{\omega_\mathrm{n}}$ accordingly.
\begin{algorithm}[!t]
\caption{Parameter Estimation from Step Response}
\label{para_algo}
\begin{algorithmic}[1]
\STATE Input: Step response $\bar{x}(t)$ \\[0.5em]
\IF {$\frac{\mathrm{d}\bar{x}(t)}{\mathrm{d}t} \neq 0, \; \forall t \in (0,\infty)$} 
\vspace{0.5em}
    \STATE $Mp = 0$, Assume $\hat{\zeta} = 1$; (critically damped case)
    \vspace{0.5em}
    \STATE Compute:
    \[
       \hat{\omega_\mathrm{n}} = \frac{4}{\hat{\zeta} T_\mathrm{s}};
    \]
\ELSIF {$\frac{\mathrm{d}\bar{x}(t)}{\mathrm{d}t} = 0, \; \forall t \in (0,\infty)$ and $\Delta \bar{x}(t_{\mathrm{p}k}) = 0$}  
\vspace{0.5em}
    \STATE $\bar{x}(t_{\mathrm{p}1}) = \bar{x}(t_{\mathrm{p}2}) = \bar{x}(t_{\mathrm{p}3}) = ...$;
\vspace{0.5em}
    \STATE Assume $\hat{\zeta} = 0$; (undamped case)
\vspace{0.5em}
    \STATE Compute:
    \[
       \hat{\omega_\mathrm{n}} = \frac{2\pi}{\Delta t};
    \]
\ELSE
    \STATE Compute:
    \begin{align*}
    M_p = \frac{\bar{x}_{\max} - \bar{x}(\infty)}{\bar{x}(\infty)}; \\[0.5em]
    \hat{\zeta} = \frac{-\ln(M_p)}{\sqrt{\pi^2 + (\ln(M_p))^2}}; \\[0.5em]
     \hat{\omega_\mathrm{n}} = \frac{\pi}{T_\mathrm{Mp}\sqrt{1-\hat{\zeta}^2}};
    \end{align*}

\ENDIF
\STATE Output: Estimated $(\hat{\zeta}, \ \hat{\omega_\mathrm{n}})$.
\end{algorithmic}
\end{algorithm}
%
%
%
\subsection{Fixed Estimation Time for Closed Form Input Shaper}
\begin{figure}
    \centering
    \includegraphics[clip, trim= 20 670 270 50,width=1\linewidth]{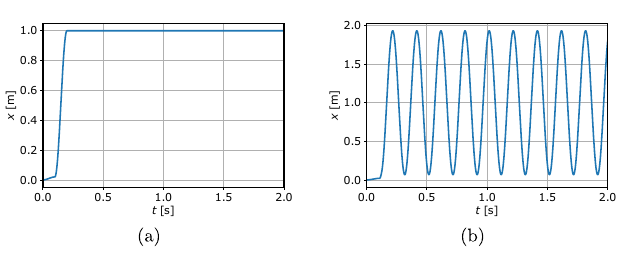}
    \caption{Reproduced results from~\cite{stein_aruco_2024} with different estimation times: (a) $\tau < T$, (b) $\tau > T$.}
    \label{fig:aruco}
\end{figure}

The fixed-time approach, as presented in~\cite{stein_aruco_2024}, modifies \eqref{eq:is_nom} as:
\begin{equation}
    G(s) = \mathcal{K} + A e^{-sT} + (1 - \mathcal{K} - A) e^{-2sT},
\end{equation}
where the switch time $T$ is treated the same as in a standard input shaper from \eqref{eq:is_nom}; hence, the estimation time $\tau$ does not explicitly appear in the formulation. Consequently, if the user selects an estimation time longer than the (unknown) switch time, the algorithm fails to achieve optimal input shaping. The reproduced results from~\cite{stein_aruco_2024} confirm that the method performs effectively when the estimation time is chosen appropriately ($\tau < T$), as shown in Fig.~\ref{fig:aruco}~(a), where the system remains at the defined position. However, performance degrades when $\tau > T$, as illustrated in Fig.~\ref{fig:aruco}~(b), where the system fails to maintain its position and exhibits sustained oscillations, resulting in ineffective vibration suppression.
%
%
%
\subsection{Arbitrary Estimation Time for Closed Form Input Shaper} \label{sec:input_shape}
This approach forms the backbone of the proposed method, where $\tau$ and $\mathcal{K}$ (as shown in Fig.~\ref{fig:block}) are explicitly incorporated into \eqref{eq:is_nom} as follows:
\begin{align} \label{eq:opt_IS}
    G(s) = \mathcal{K} + Ae^{-s(\tau + T)} + (1-A-\mathcal{K})e^{-s(\tau+2T)},
\end{align}
where $A$ the magnitude of the first impulse at $t = \tau + T$, and $(1-A-\mathcal{K})$ the magnitude of the second impulse at \mbox{$t = \tau + 2T$}. Since the switch time applies only after the estimation phase, the performance is independent of the chosen $\tau$. In this formulation, $\mathcal{K}$ and $\tau$ are user-defined, leaving $A$ and $T$ as the only unknowns. These values are obtained by substituting \mbox{$s = \pm j\omega_\mathrm{n}$} for the undamped case and \mbox{$s = -\zeta\omega_\mathrm{n} \pm j\omega_\mathrm{n}\sqrt{1-\zeta^2}$} for the damped case. 1) An analytical solution is applied to the undamped system, while 2) a numerical procedure is used for the damped system, as detailed in the following sections.
%
%
%
\subsubsection{Analytical Approach for Undamped System}
Using \mbox{$s=\pm j \omega_\mathrm{n}$} in \eqref{eq:opt_IS} (consider only the positive root) and equating for the real and imaginary component leads to:
\begin{subequations}  
    \begin{align}
        \mathcal{K} + \cos(\omega_\mathrm{n}(2T + \tau))(1 - A - \mathcal{K}) + A\cos(\omega_\mathrm{n}(T + \tau)) = 0, \label{eq:constraint_real} \\
        -\sin(\omega_\mathrm{n}(2T + \tau))(1 - A - \mathcal{K}) - A\sin(\omega_\mathrm{n}(T + \tau)) = 0.
        \label{eq:constraint_imag}
    \end{align}
\end{subequations}
This can be simplified to the following equation:
\begin{align}
    -\sin(\omega_\mathrm{n} T) + \mathcal{K} \sin(\omega_\mathrm{n}(T + \tau)) + \mathcal{K} \sin(\omega_\mathrm{n} T) \nonumber\\
    - \mathcal{K} \sin(\omega_\mathrm{n}(2T + \tau)) = 0. 
\end{align}
Grouping terms results in:
\begin{align}
    (-1 + \mathcal{K})\sin(\omega_\mathrm{n} T) + \mathcal{K} \sin(\omega_\mathrm{n}(T + \tau)) \nonumber\\
    - \mathcal{K} \sin(\omega_\mathrm{n}(2T + \tau)) = 0. \label{eq:TDF_transcendental}
\end{align}
Now let us define the Weierstrass substitution as:
\begin{align}
    \phi = \omega_\mathrm{n}\tau,\:\:\:\theta = \omega_\mathrm{n} T;\:\:\: \beta = \tan\left(\frac{\theta}{2}\right). \label{eq:Weierstrass}
\end{align}
Then, the following trigonometric identities can be used:
\begin{subequations}
    \begin{align}
        \sin(\theta) = \frac{2\beta}{1 + \beta^2}, \\
        \cos(\theta) = \frac{1 - \beta^2}{1 + \beta^2}, \\
        \sin(\theta + \phi) = \sin(\theta)\cos(\phi) + \cos(\theta)\sin(\phi), \\
                            = \frac{2\beta \cos(\phi) + (1 - \beta^2)\sin(\phi)}{1 + \beta^2}. 
    \end{align}
    \label{eq:TDF_Weierstrass_single_angle}
\end{subequations}
Similarly, compute:
\begin{subequations}
    \begin{align}
        \sin(2\theta) = \frac{4\beta(1 - \beta^2)}{(1 + \beta^2)^2}, \\
        \cos(2\theta) = \frac{1 - 6\beta^2 + \beta^4}{(1 + \beta^2)^2}, \\
        \sin(2\theta + \phi) = \sin(2\theta)\cos(\phi) + \cos(2\theta)\sin(\phi), \\
        = \frac{4\beta(1 - \beta^2)\cos(\phi) + (1 - 6\beta^2 + \beta^4)\sin(\phi)}{(1 + \beta^2)^2}. 
    \end{align}
    \label{eq:TDF_Weierstrass_double_angle}
\end{subequations}
Substituting Eqs.~\eqref{eq:Weierstrass}-\eqref{eq:TDF_Weierstrass_double_angle} into~\eqref{eq:TDF_transcendental} results in:
\begin{align}
    (-1 + \mathcal{K}) \frac{2\beta}{1 + \beta^2}
    + \mathcal{K} \frac{2\beta \cos(\phi) + (1 - \beta^2)\sin(\phi)}{1 + \beta^2} \nonumber \\
    - \mathcal{K} \frac{4\beta(1 - \beta^2)\cos(\phi) + (1 - 6\beta^2 + \beta^4)\sin(\phi)}{(1 + \beta^2)^2} = 0.
\end{align}
Multiply through by $(1 + \beta^2)^2$ to eliminate denominators leads to:
\begin{align}
    (-1 + \mathcal{K}) 2\beta(1 + \beta^2) \nonumber\\
    + \mathcal{K} \left[2\beta \cos(\phi)(1 + \beta^2) + (1 - \beta^2)\sin(\phi)(1 + \beta^2) \right] \nonumber\\
    - \mathcal{K} \left[4\beta(1 - \beta^2)\cos(\phi) + (1 - 6\beta^2 + \beta^4)\sin(\phi) \right] = 0, \label{eq:TDF_polynomial}
\end{align}
which can be further simplified to:
\begin{align}
    2 \beta\left(\mathcal{K} - \mathcal{K}\cos(\phi) + \mathcal{K}\beta^2 - \beta^2 + 3\mathcal{K} \beta \sin(\phi) \right. \nonumber\\
    \left. + 3\mathcal{K}\beta^2\cos(\phi) - \mathcal{K}\beta^3\sin(\phi) - 1\right) = 0. \label{eq:TDF_polynomial_simplified}
\end{align}
The roots of the system are then given as:
\begin{subequations}
    \begin{align}
        \beta_1 = 0, \\
        \beta_2 = \mathcal{A} + \frac{\mathcal{B}}{\mathcal{U}} + \mathcal{U}, \\
        \beta_3 = \mathcal{A} - \frac{1}{2}\mathcal{U} - \frac{\mathcal{B}}{2\mathcal{U}} - i \frac{\sqrt{3}}{2}\left(\mathcal{U} - \frac{\mathcal{B}}{\mathcal{U}}\right), \\
        \beta_4 = \mathcal{A} - \frac{1}{2}\mathcal{U} - \frac{\mathcal{B}}{2\mathcal{U}} + i \frac{\sqrt{3}}{2}\left(\mathcal{U} - \frac{\mathcal{B}}{\mathcal{U}}\right),
    \end{align}
\end{subequations}
where,
\begin{subequations}
    \begin{align}
        \mathcal{A} = \frac{\mathcal{K} + 3\mathcal{K} \cos \phi - 1}{3 \mathcal{K} \sin \phi}, \\
        \mathcal{B} = \frac{(\mathcal{K} + 3\mathcal{K} \cos \phi - 1)^2}{9 \mathcal{K}^2 \sin^2 \phi} + 1, \\
        \mathcal{C} = \frac{\mathcal{K} + 3\mathcal{K} \cos \phi - 1}{2 \mathcal{K} \sin \phi} - \frac{\mathcal{K} \cos \phi - \mathcal{K} + 1}{2 \mathcal{K} \sin \phi} \nonumber \\
        + \frac{(\mathcal{K} + 3\mathcal{K} \cos \phi - 1)^3}{27 \mathcal{K}^3 \sin^3 \phi}, \\
        \mathcal{D} = \sqrt{\mathcal{C}^2 - \mathcal{B}^3}, \quad \mathcal{U} = (\mathcal{C} + \mathcal{D})^{1/3}.
    \end{align}
\end{subequations}
Now~\eqref{eq:TDF_polynomial_simplified} can be solved for $\beta$ based on $\mathcal{K}$, $\omega_\mathrm{n}$, and $\tau$ in closed-form. The corresponding switch times are then expressed as:
\begin{align} \label{eq:analytical_1}
    T_{k}^{(i)} = \frac{2}{\omega_\mathrm{n}} \arctan(\beta_i) + \frac{2\pi k}{\omega_\mathrm{n}},
\end{align}
where $i\in[1,2,3,4]$ and $k \in \mathbb{Z}$. Initially, all four solutions of $\beta$ are considered. Once the parameter values for $\mathcal{K}, \ \omega_\mathrm{n}$, and $\tau$ are specified, only the purely real roots are retained. The coefficient $A$ is then calculated as:
\begin{align} \label{eq:analytical_2}
    A = -\frac{\mathcal{K} - \cos(\omega_\mathrm{n} (\tau + 2T))(\mathcal{K} - 1)}
{\cos(\omega_\mathrm{n} (\tau + T)) - \cos(\omega_\mathrm{n} (\tau + 2T ))}.
\end{align}
%
%
%
\subsubsection{Numerical Approach for Damped System}
By substituting $s = -\zeta\omega_\mathrm{n} \pm j\omega_\mathrm{n}\sqrt{1-\zeta^2}$ into \eqref{eq:opt_IS} (consider only the positive root) and setting both the real and imaginary parts to zero, we obtain:
\begin{subequations}
    \begin{align}
    \mathcal{K} + A e^{\zeta \omega_\mathrm{n} (\tau+T)} 
    \cos(\omega_\mathrm{n} \sqrt{1-\zeta^2}\ (\tau+T)) + \nonumber \\
    (1 - \mathcal{K} - A) e^{\zeta \omega_\mathrm{n} (\tau+2T)} 
    \cos(\omega_\mathrm{n} \sqrt{1-\zeta^2}\ (\tau+2T)) = 0,
    \label{eq:num_damped1}\\
    A \ e^{\zeta \omega_\mathrm{n} (\tau + T)} \ 
    \sin(\omega_\mathrm{n} \sqrt{1 - \zeta^2} (\tau + T)) + \nonumber \\ (1 - \mathcal{K} - A)\
    e^{\zeta \omega_\mathrm{n} (\tau + 2T)} \ 
    \sin(\omega_\mathrm{n} \sqrt{1 - \zeta^2} (\tau + 2T)) = 0. \label{eq:num_damped2}
    \end{align}
\end{subequations}
Since both exponential and sinusoidal terms appear in \eqref{eq:num_damped1} and \eqref{eq:num_damped2}, analytical simplification is impractical. Therefore, a numerical solution is employed, since the problem reduces to two equations with two unknowns solving for $A$ and $T$.
%
%
%
\section{RESULTS}
\label{sec:results}
Since a black-box second-order system is considered, the scenarios in this work include $\zeta \in [0,1]$ and $\omega_\mathrm{n} \in [\pi, 3000\pi]$. The proposed method is applied to representative cases across this range, and the corresponding results are presented in Fig.~\ref{fig:all_results}.
%
%
%
\subsection{Performance of the Proposed Method Across Diverse Scenarios} \label{sec:result_A}
\begin{figure*}
    \centering
    \includegraphics[clip, trim= 20 425 20 20,width=\linewidth]{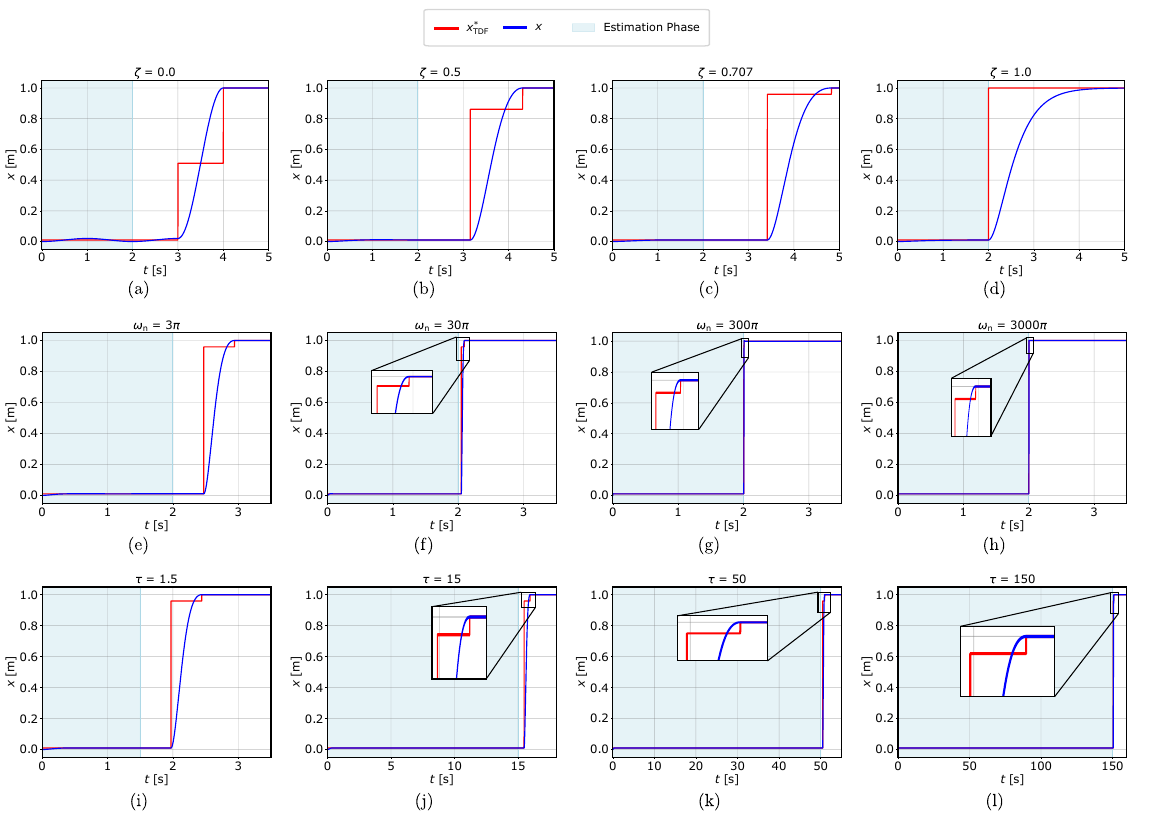}
    \caption{Performance of the proposed method, 
    (a-d) varying $\zeta$ with $\tau = 2\mathrm{s}$ and $\omega_\mathrm{n} = \pi \ \mathrm{rad/s}$,
    (e-h) varying $\omega_\mathrm{n}$ with $\tau = 2\mathrm{s}$ and $\zeta = 0.707$,
    (i-l) varying $\tau$ with $\omega_\mathrm{n} = 3\pi \ \mathrm{rad/s}$ and $\zeta = 0.707$.}
    \label{fig:all_results}
\end{figure*}
In all cases, a unit step input is applied, and as shown in Fig.~\ref{fig:all_results}, the feedforward control performs consistently across a range of black-box second-order systems. Undamped, damped, and critically damped cases are illustrated in Fig.~\ref{fig:all_results}(a–d), where the system responses exhibit no oscillation or overshoot. Although the responses appear slower due to the TDF reference, this effect is primarily related to the system’s natural frequency. This is confirmed in Fig.~\ref{fig:all_results}(e–h), where higher $\omega_\mathrm{n}$ values yield responses that closely match the TDF reference. 

The proposed method is further evaluated under varying estimation times. Unlike the approach in \cite{stein_aruco_2024}, the results in Fig.~\ref{fig:all_results}(i–l) demonstrate that the system maintains excellent performance regardless of estimation duration. Overall, the responses show no overshoot or oscillation across damped or undamped, slow or fast systems, or under different estimation times. For real-world applications such as gantry crane motion or 3D printer nozzle positioning, this translates into minimal or no vibration, effectively demonstrating vibration suppression.

Moreover, the numerical performance of the estimator (Algorithm~\ref{para_algo}) is summarized in Table~\ref{tab:num_result}, showing close agreement with the true system parameters. The rows are grouped according to Figs.~\ref{fig:all_results}(a–d), (e–h), and (i–l), with varying scenarios highlighted in light blue for clarity and discrepancies highlighted in light red. Small discrepancies between the true and estimated values can be observed in extreme cases, such as critically damped systems or very high natural frequency. These discrepancies arise from the assumptions made in Algorithm~\ref{para_algo} for the critically damped case, as well as from its difficulty in capturing $T_\mathrm{Mp}$ and $T_\mathrm{s}$ in very fast systems. This issue can be mitigated by reducing the time step (i.e., increasing the resolution), though at the cost of higher computational effort. Nevertheless, such deviations are not critical, as the feedforward control performance remains consistent with expectations, as shown in Fig.~\ref{fig:all_results}.
\begin{table}[htbp]
\caption{Performance of the parameter estimation (Algorithm~\ref{para_algo}) under parameter sweeps of $\zeta$, $\omega_\mathrm{n}$, and $\tau$.}
\label{tab:num_result}
\renewcommand{\arraystretch}{1.5}
\centering
\begin{tabular}{l l l l l}
\Xhline{2.5\arrayrulewidth} 
 $\zeta$ & $\hat{\zeta}$ & $\omega_{\mathrm{n}} (\mathrm{rad/s})$ & $\hat{\omega_{\mathrm{n}}} (\mathrm{rad/s})$ & $\tau$ (s) \\
\Xhline{2.5\arrayrulewidth} 
\cellcolor{lightblue}0.0   & $4.2 \cdot 10^{-16}$ & $\pi$        & $\approx \pi$        & 2 \\
\cellcolor{lightblue}0.5   & 0.5                  & $\pi$        & $\approx \pi$        & 2 \\
\cellcolor{lightblue}0.707 & 0.707                & $\pi$        & $\approx \pi$        & 2 \\
\cellcolor{lightblue}1.0   & 1.0                  & $\pi$        & \cellcolor{lightred}$\approx 0.7 \pi$ & 2 \\
\Xhline{2.5\arrayrulewidth} 
0.707 & 0.707                & \cellcolor{lightblue}$3 \pi$    & $\approx 3 \pi$      & 2 \\
0.707 & 0.707                & \cellcolor{lightblue}$30 \pi$   & $\approx 30 \pi$     & 2 \\
0.707 & \cellcolor{lightred}0.7106 & \cellcolor{lightblue}$300 \pi$  & \cellcolor{lightred}$\approx 285 \pi$  & 2 \\
0.707 & \cellcolor{lightred}0.7106 & \cellcolor{lightblue}$3000 \pi$ & \cellcolor{lightred}$\approx 2850 \pi$ & 2 \\
\Xhline{2.5\arrayrulewidth} 
0.707 & 0.707 & $3 \pi$ & $\approx 3 \pi$ & \cellcolor{lightblue}1.5 \\
0.707 & 0.707 & $3 \pi$ & $\approx 3 \pi$ & \cellcolor{lightblue}15 \\
0.707 & 0.707 & $3 \pi$ & $\approx 3 \pi$ & \cellcolor{lightblue}50 \\
0.707 & 0.707 & $3 \pi$ & $\approx 3 \pi$ & \cellcolor{lightblue}150 \\
\Xhline{2.5\arrayrulewidth} 
\end{tabular}
\end{table}
%
%
%
\subsection{Performance of the Optimal Input Shaper under Multiple Step-Wise Reference Changes}
\begin{figure}
    \centering
    \includegraphics[clip, trim= 30 585 305 5,width=0.95\linewidth]{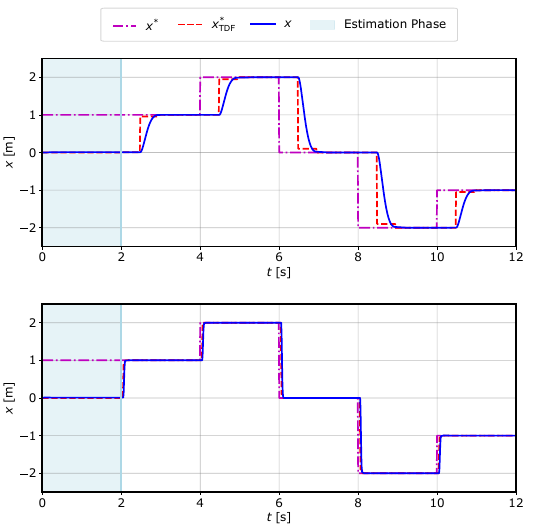}
    \caption{Feedforward control for step-wise reference tracking: (Top) $\omega_\mathrm{n} = 3\pi$ rad/s; (Bottom) $\omega_\mathrm{n} = 30\pi$ rad/s.}
    \label{fig:step_wise}
\end{figure}
To further evaluate the proposed method, step-wise reference changes around positive, negative, and zero values are applied instead of a single unit step. As discussed in Algorithm~\ref{para_algo}, the system shows little response to the reference during the user-defined estimation phase, as illustrated in Fig.~\ref{fig:step_wise}. After estimation, the input $x^{*}$ is shaped into the time-delay reference $x_{\mathrm{TDF}}^{*}$, and the system begins tracking the TDF reference, showing consistent performance with the results in Fig.~\ref{fig:all_results} and exhibiting vibration-free responses regardless of the number of step-wise changes. In Fig.~\ref{fig:step_wise} (top), an apparent delay is observed due to the TDF; however, this effect is primarily attributed to the system’s natural frequency, as discussed in Sec.~\ref{sec:result_A}. In contrast, Fig.~\ref{fig:step_wise} (bottom) shows that for a fast system, the response closely matches the reference with negligible delay.
%
%
%
\subsection{Dependency of Optimal Input Shaper Design on System Parameters}
\begin{figure}
    \centering
     \includegraphics[clip, trim= 20 700 265 20,width=\linewidth]{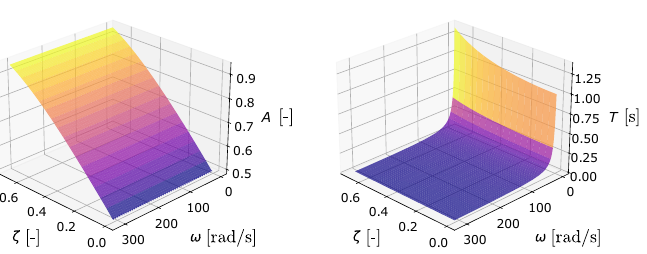}
    \caption{Dependency of input shaper parameters $A$ and $T$ on $\zeta$ and $\omega_\mathrm{n}$ under parameter sweep.}
    \label{fig:3D_AT}
\end{figure}
As seen in \eqref{eq:analytical_1}, \eqref{eq:analytical_2}, \eqref{eq:num_damped1} and \eqref{eq:num_damped2}, the TDF parameters $A$ and $T$ depend on the user-defined values $\mathcal{K}$ and $\tau$, as well as the system parameters $\zeta$ and $\omega_\mathrm{n}$. Since the user-defined values cannot be controlled, parameter sweeps are conducted over $\zeta$ and $\omega_\mathrm{n}$ to evaluate their influence on $A$ and $T$, as illustrated in Fig.~\ref{fig:3D_AT}. The results indicate that $A$ is primarily affected by $\zeta$, while $T$ is strongly dependent on $\omega_\mathrm{n}$. This suggests the feasibility of developing a simple linear regression model, enabling the input-shaper block in Fig.~\ref{fig:block} to be replaced with a trained regression model as an alternative when online optimal input shaping is not available.
%
%
%
\subsection{Key Consideration of the Proposed Method}
An important aspect of the proposed method is that the estimation time is user-defined and must be chosen realistically. For slow systems, the estimation time should be sufficiently long to capture at least one full cycle of the natural response or to extend beyond the transient period. In contrast, for systems with high natural frequencies, a shorter estimation time is sufficient to capture the system response accurately.
%
%
%
\section{CONCLUSIONS}
\label{sec:conclusion}
The manuscript presents a method that performs online parameter estimation for a black-box second-order system and subsequently designs an input shaper, ensuring that the system output follows the reference trajectory without vibration or overshoot. As demonstrated in Sec.~\ref{sec:results}, the approach was effective across all tested cases, with $\omega_\mathrm{n}$ up to $3000\pi$, and it is expected to generalize to much higher frequencies. Furthermore, this work highlights the limitations of the fixed estimation-time approach in \cite{stein_aruco_2024} and improves upon it, by letting the user choose an arbitrary estimation time for the shaper design. Additionally, a closed-form solution is provided for the undamped case which enables a very fast input shaper design. As future work, the online design of the optimal input shaper may be replaced with a trained model to obtain $x_{\mathrm{TDF}}^*$ more efficiently. This study serves as a preliminary step toward extending the proposed framework to general black-box systems beyond a second-order single mode case.
\bibliographystyle{IEEEtran}
\bibliography{references}
\addtolength{\textheight}{-12cm}
\end{document}